\documentclass[aps,prl,twocolumn,superscriptaddress,groupedaddress]{revtex4}   
\usepackage{graphicx}  
\usepackage{dcolumn}   
\usepackage{bm}        
\usepackage{amssymb}   
\usepackage{hyperref}
\usepackage{xcolor}
\hypersetup{colorlinks = true,linkcolor = blue,citecolor= blue}
\begin{document}

\title{Gravitational waves from oscillons after inflation}

\author{Stefan Antusch}
\affiliation{Deparment of Physics, University of Basel, Klingelbergstr. 82, CH-4056 Basel, Switzerland}
\affiliation{Max-Planck-Institut f\"ur Physik (Werner-Heisenberg-Institut),
F\"ohringer Ring 6, D-80805 M\"unchen, Germany}

\author{Francesco Cefal\`a}
\affiliation{Deparment of Physics, University of Basel, Klingelbergstr. 82, CH-4056 Basel, Switzerland}

\author{Stefano Orani}
\affiliation{Deparment of Physics, University of Basel, Klingelbergstr. 82, CH-4056 Basel, Switzerland}

\date{\today}

\begin{abstract}
We investigate the production of gravitational waves during preheating after inflation in the common case of field potentials that are asymmetric around the minimum. In particular, we study the impact of oscillons, comparatively long lived and spatially localized regions where a scalar field (e.g.\ the inflaton) oscillates with large amplitude. Contrary to a previous study, which considered a symmetric potential, we find that oscillons in asymmetric potentials associated with a phase transition can generate a pronounced peak in the spectrum of gravitational waves, that largely exceeds the linear preheating spectrum. We discuss the possible implications of this enhanced amplitude of gravitational waves. For instance, for low scale inflation models, the contribution from the oscillons can strongly enhance the observation prospects at current and future gravitational wave detectors.

\end{abstract}

\maketitle

{\bf Introduction:} Inflation is a very successful paradigm for early universe cosmology. The accelerated expansion can solve the horizon and flatness problems, while the quantum fluctuations of the inflaton field provide the seed for structure in the universe.   
After inflation, the potential energy of the inflaton is transferred to a thermal bath of the matter species present in the universe today in a process called reheating. The early stage of reheating, referred to as preheating, is often governed by non-linear dynamics of the inflaton field and other fields coupled to it, typically resulting in inhomogeneous field configurations. A generic consequence of preheating is the production of a stochastic background of gravitational waves (GW)  \cite{Khlebnikov:1997di,Easther:2006gt}. 

Observations of the CMB \cite{Ade:2015lrj} point to adiabatic, nearly Gaussian primordial fluctuations as predicted by simple one-field slow-roll models of inflation. Furthermore, tight constraints on the ratio of tensor to scalar fluctuations, $r<0.09$ at $95\%$ C.L.\ \cite{Ade:2015lrj}, can be seen as a hint toward small-field models of inflation taking place below the Planck scale. The red-tilted spectral index, $n_{\rm s} =0.968\pm0.006$ at $68\%$ C.L.\ \cite{Ade:2015lrj}, then points to a negatively curved inflaton potential, where inflation happens along a ``plateau'' with large potential energy, i.e.\ along a flat ``hilltop'' \cite{Martin:2013nzq,Linde:1981mu}. Such inflaton potentials are also attractive because they appear in particle physics models where a phase transition at high energies takes place (see e.g.\ \cite{Linde:1981mu,Antusch:2008gw}). These potentials are in general asymmetric around the minimum.

Reheating in these models generically features oscillons, comparatively long lived and spatially localized regions where the inflaton oscillates with large amplitude. Oscillons can be produced during preheating after different models of inflation  \cite{Copeland:2002ku,Broadhead:2005hn,Amin:2011hj,Gleiser:2014ipa,Antusch:2015nla}, as well as in various types of field theories \cite{Gleiser:1993pt,Copeland:1995fq,Farhi:2005rz,Fodor:2006zs,Graham:2006vy,Gleiser:2007te,Achilleos:2013zpa,Bond:2015zfa}. In \cite{Amin:2013ika}, it has been shown that they form when a scalar field oscillates in a potential that opens up away from the minimum, i.e.\ that is shallower than quadratic. The hilltop potentials mentioned above have this property on one side of the minimum, while on the other side they are steeper. Nevertheless, oscillons are a characteristic feature of the reheating dynamics of this class of models. 
Despite the fact that the potential is steeper than quadratic on one side, the oscillons are ``long-lived'' and can survive at least several $e$-folds after the end of inflation \cite{Antusch:2015ziz,Graham:2006xs}, potentially contributing to the production of GW. Interactions with other fields can affect the oscillons in some cases, e.g.\ when a parametric resonance occurs, however in general they do not have a significant impact during the first few $e$-folds of reheating (see e.g.\ \cite{Antusch:2015ziz}).

So far, effects of oscillons on the production of GW have been studied in \cite{Zhou:2013tsa} in the context of axion monodromy inflation \cite{Silverstein:2008sg}, a large-field model that is symmetric around the minimum.
It was found that oscillons contribute to GW production when they form after inflation, generating a small peak in the GW spectrum. However, in that model the oscillons quickly become spherically symmetric, suppressing the production of GW. As a consequence, the GW peak stops growing very soon, until the oscillons eventually decay. Their decay, which was not studied in \cite{Zhou:2013tsa}, is another potential source of GW. 

In this letter, we study GW production from oscillons in field potentials which are asymmetric around the minimum, as typical in plateau inflation or hilltop inflation models embedded into high energy particle physics. We find that oscillons in such asymmetric potentials converge less efficiently to a spherical shape and GW production continues long after the oscillon formation phase. As a result, the GW spectrum continues growing during the ``oscillon phase'', i.e.\ the phase after oscillon formation and before they decay. This continuous growth can yield a pronounced peak in the GW spectrum which largely exceeds the GW from linear preheating. We argue that this is a generic effect in asymmetric potentials, and discuss possible implications of the enhanced GW signal. 

{\bf Framework:} 
As mentioned above, models of hilltop inflation are favoured by recent CMB observations and offer attractive links to particle physics phase transitions. We choose a simple realisation of such potentials, of the form  
\begin{equation}
V(\phi) \,=\,V_0\,\left(1-\frac{\phi^p}{v^p}\right)^2\,,
\label{eq:hilltop}
\end{equation}
where $V_0$ is the potential energy on top of the hill, $p \ge 3$, and $\phi$ is a real scalar field with $|\phi| =  v$ holding at the minimum of the potential. For example, $\phi$ can be identified with an order parameter of a second order phase transition, where some symmetry gets spontaneously broken. 
The universe inflates while $\phi$ rolls away from the maximum at $\phi=0$ and inflation ends when the curvature of the potential becomes too large and the inflaton accelerates toward $v$. In this model, $V_0$ is fixed by the amplitude of the primordial curvature perturbation $A_{\rm s} \simeq 2.2\times10^{-9}$ \cite{Ade:2015lrj}. 
For $p=6$ and $v=10^{-2} m_{\rm Pl}$, which we will use as example in this study, we have  
$n_{\rm s} \simeq 0.96$, $r\simeq 10^{-12}$ and 
$V_0=24\pi^2\varepsilon A_{\rm s}m^4_{\rm Pl}\simeq10^{-13}v^3m_{\rm Pl}\simeq10^{-19}m_{\rm Pl}^4$
with the slow-roll parameter $\varepsilon\equiv\frac{1}{2}m^2_{\rm Pl}(\partial V/\partial\phi)^2/V^2$ 
evaluated $N\simeq60$ $e$-folds before the end of inflation.
 
Around the minimum at $\phi = v$, the potential is highly asymmetric, 
with an inflection point toward the plateau for $\phi<v$ and steeper than quadratic for $\phi>v$. Thus, such potentials support oscillons only on one side, $\phi<v$. As mentioned above, oscillons in this type of potential form after inflation \cite{Gleiser:2014ipa,Antusch:2015nla}, when the inflaton accelerates toward the minimum and undergoes a series of tachyonic oscillations, periodically crossing the inflection point at $\phi<v$. These oscillons are then separated by a characteristic distance related to the frequency of the tachyonic oscillations, which is proportional to the mass of the inflaton around the minimum $m_\phi \propto \sqrt{V_0}/v$. 

The above scenario is very minimal, and ties $V_0$ to the amplitude of the curvature perturbation $A_{\rm s}$ once $v$ is fixed. For $v\simeq 10^{-2} m_{\rm Pl}$ leading to $V_0\simeq10^{-19}m_{\rm Pl}^4\simeq \mathcal{O}(10^{13}\,{\rm GeV})^4$, this also fixes today's frequency of the GW generated during preheating to $f\simeq10^{10}\,{\rm Hz}$, many orders of magnitude beyond the frequencies that can be reached by currently envisaged experiments. 
Lower frequencies in the observable range are possible when the scenario of Eq.~(\ref{eq:hilltop}) is generalized. For instance, $\phi$ does not necessarily have to be the inflaton field itself. 

Very similar (p)reheating and oscillons can indeed emerge in scenarios where a second field acts as the inflaton, i.e.\ in hybrid-like inflation models. The potentials of these models have the form
 \begin{equation}
V(\chi,\phi) \,=\,V_0\,\left(1-\frac{\phi^p}{v^p}\right)^2 + V_{\rm inf}(\chi,\phi)\,,
\label{eq:tribrid}
\end{equation}
where now $p\ge 2$ and $V_{\rm inf}(\chi,\phi)$ is responsible for the $N\sim60$ $e$-folds of inflation, with $\phi \approx 0$ after inflation \footnote{To avoid the production of topological defects after inflation
a slight deformation of the potential will be considered, e.g.\ by replacing $V_0  \rightarrow V_0\,(1+\beta\,\phi)$. The parameter $\beta$ drives the inflationary trajectory to $\phi \neq 0$ and also accelerates the transition to $\phi=v$ at the end of inflation, thus preventing a second phase of inflation when $p\geq3$.}. The choice $p=2$ includes the case of hybrid inflation \cite{Linde:1993cn}, for which the GW signal has been studied e.g.\ in \cite{Dufaux:2008dn} \footnote{Oscillons have been observed in hybrid inflation (see e.g.\ \cite{Copeland:2002ku,Broadhead:2005hn}), although their contribution to the GW spectrum needs further study. In \cite{Dufaux:2008dn}, the authors calculated the GW spectrum produced during preheating after hybrid inflation but did not discuss oscillons and did not observe the pronounced oscillon peak.}. 
Furthermore, we may also consider $p\ge3$ as e.g.\ in the tribrid inflation models of \cite{Antusch:2012jc}, which would then give preheating dynamics analogous to model (\ref{eq:hilltop}).
Oscillons in this scenario form after inflation during (p)reheating when $\phi$ is rolling toward the minimum of the potential at $\phi = v$, as discussed above. 

The main difference between the models (\ref{eq:hilltop}) and (\ref{eq:tribrid}) is that in (\ref{eq:tribrid}), $V_0$ and $v$ have become essentially free parameters, which opens up the possibility to realize a low-scale phase transition (with e.g.\ $V_0 \sim {\cal O}({\rm 100 \:TeV})^4$) such that the frequency of the GW lies in the observable range of present and future experiments. 
Furthermore we note that potentials of the form of Eq.\ (\ref{eq:hilltop}) can also arise in particle physics models with phase transitions independent of inflation, and in this case $V_0$ may also lie in the ${\cal O}({\rm 100 \:TeV})^4$ range.

{\bf GW spectrum from lattice simulations:}
We have simulated the production of GW during preheating in the models (\ref{eq:hilltop}) and (\ref{eq:tribrid}) using 3-dimensional lattice simulations. To this end, we used a modified version of LATTICEEASY \cite{Felder:2000hq}. 
For further discussion of GW production in lattice simulations of preheating, see e.g.\ \cite{Dufaux:2007pt,GarciaBellido:2007af}.

The original version of the program solves a discretized version of the non-linear scalar field dynamics in a Friedmann-Lema\^itre-Robertson-Walker universe (FLRW). We consider a real scalar field $\phi$ and solve the following set of equations in a portion of comoving volume $\mathcal{V}$:
\begin{eqnarray}
\ddot{\phi}\, + \, 3H\dot{\phi}\, - \,\frac{1}{a^2}\nabla^2\phi\, + \,\frac{\partial V}{\partial\phi} \, = \, 0\,,\nonumber\\
H^2\, = \,\frac{1}{3m^2_{\rm{Pl}}}\left\langle V\,  + \,\frac{1}{2}\dot{\phi}^2\, + \frac{1}{2a^2}\left|\nabla\phi\right|^2 \right\rangle_{\mathcal{V}}\,,
\label{eq:EOM_fld}
\end{eqnarray}
where $\langle ... \rangle_{\mathcal{V}}$ denotes a spatial average over $\mathcal{V}$.
Furthermore, we have implemented additional code that allows to simultaneously solve the equations of motion of GW. They correspond to the transverse-traceless (TT) part $h_{ij}$ of the tensor perturbations of flat FLRW. In the synchronous gauge, the line element can be written as 
\begin{equation}
ds^2\, = \, -dt^2\, + \,a^2(t)(\delta_{ij}\, + \,h_{ij})dx^idx^j \,,
\end{equation}
with $\partial_ih_{ij}=h_{ii}=0$. The equations of motion are
\begin{equation}
\ddot{h}_{ij}\, + \,3H\dot{h}_{ij}\, - \,\frac{1}{a^2}\nabla^2h_{ij}\, = \,\frac{2}{m^2_{\rm Pl}a^2}\Pi^{\rm TT}_{ij}\,,
\label{eq:EOM_tensor}
\end{equation}
where $\Pi^{\rm TT}_{ij}\,=\,[\partial_i\phi\partial_j\phi]^{\rm TT}$ is the TT part of the anisotropic stress tensor (for more details see e.g.\ \cite{Figueroa:2011ye}).

The GW energy density is then given by 
\begin{equation}
\rho_{\rm GW}(t)\, = \, \frac{m^2_{\rm Pl}}{4}\left\langle \dot{h}_{ij}(\mathbf{x},t)\dot{h}_{ij}(\mathbf{x},t)\right\rangle_{\mathcal{V}}\,.
\label{eq:rho_GW}
\end{equation}
The spectrum of the energy of GW per logarithmic momentum interval observable today, and its frequency, are

\begin{eqnarray}
\Omega_{\rm GW,\,0}h^2 &=&  \left.\frac{h^2}{\rho_c}k\frac{d\,\rho_{\rm GW}}{d \, k}\right|_{t_0} = \left.\frac{h^2}{\rho_c}k\frac{d\,\rho_{\rm GW}}{d \, k}\right|_{t_e}\frac{a_e^4\rho_e}{a_0^4\,\rho_{c,\,0}}\,\nonumber\\
&=& \frac{4.3}{10^{5}}\Omega_{\rm GW,\,e}\left(\frac{a_e}{a_*}\right)^{1-3w}\left(\frac{g_*}{g_0}\right)^{-1/3}\,,
\label{eq:OmegaGW}\\
f &=& \frac{k}{a_e\rho_e^{1/4}}\left(\frac{a_e}{a_*}\right)^{\frac{1-3w}{4}}4\times10^{10}\:{\rm Hz}\,,
\label{eq:fGW}
\end{eqnarray}
where, respectively, the subscript $0$ indicates quantities evaluated today, $e$ at the end of the lattice simulations and $*$ at the end of reheating, while $\rho_{c,\,0}$ is the critical energy density today, $g$ the number of light degrees of freedom and $w$ is the mean equation of state between $t_e$ and $t_*$ (see e.g.\ \cite{Dufaux:2007pt} for more details). In our calculations we use $g_*/g_0=100$.   

In order to study the production of GW during preheating in the models of Eqs.~(\ref{eq:hilltop}) and (\ref{eq:tribrid}) we performed $3$-dimensional lattice simulations with $128^3$ points in a box with comoving volume $\mathcal{V}\equiv L^3\sim(0.01/H_{i})^3$, where $H_{i}$ is the Hubble parameter at the beginning of the simulations and the initial scale factor $a_{i}=1$. 
The parameters and setup of the lattice simulations are given in Table \ref{tab:IC_and_parameters}. 
The fluctuations of the field and its derivative were initialized using the usual prescription \cite{Polarski:1995jg,Khlebnikov:1996mc}, i.e.\ as stochastic variables with a variance reproducing the two-point function of the quantum vacuum fluctuations. For details on the numerical implementation, see \cite{Felder:2000hq}. The tensor perturbations and their derivatives were initialized as zero.
We stopped the simulations about $3$ $e$-folds after the end of inflation, before fluctuations on the smallest distances of the lattice become important and the simulations break down.

\begin{table}[h]
\caption{\label{tab:IC_and_parameters} Initial conditions and parameters of simulations of the models of Eqs.~(\ref{eq:hilltop}) and (\ref{eq:tribrid}) on a 
$3$-dimensional lattice with $128^3$ points.}
\begin{ruledtabular}
\begin{tabular}{c|c|c|c|c|c|c}
Model &$L\,H_{i}$ & $v/m_{\rm Pl}$ &
$V_0/m^4_{\rm Pl}$ &$p$&$\langle \phi_{i} \rangle/v$ & $\langle \dot{\phi}_{i} \rangle/v^{2}$  \\
\hline
Eq.~(\ref{eq:hilltop})$\vphantom{\frac{1^2}{2}}$& 0.01 &  $10^{-2}$  & $10^{-19}$
& 6 & 0.08 & $2.49\times10^{-9}$  \\
Eq.~(\ref{eq:tribrid})& 0.01 & $10^{-2}$ & free
& 6 &  0 &  0\\
\end{tabular}
\end{ruledtabular}
\end{table}

We note that regarding the produced GW spectrum, we did not find any noticeable difference between the models (\ref{eq:hilltop}) and (\ref{eq:tribrid}): In particular, for model (\ref{eq:tribrid}), we considered a tribrid inflation scenario with $p=6$ and with a linear deformation of the potential $V_0 \rightarrow V_0\,(1+\beta\,\phi)$ (see e.g.\ \cite{Antusch:2013toa}) and convinced ourselves that the deformation has no impact on the first $\sim3$ $e$-folds after inflation if $\beta m_{\rm Pl}\lesssim \sqrt{2}$. For the simulation we used $\beta m_{\rm Pl} = \sqrt{2}$. 
Interestingly, since the equations of motion Eqs.~(\ref{eq:EOM_fld}) and (\ref{eq:EOM_tensor}) are invariant under a simultaneous rescaling of the potential and of distance and time units, the GW amplitude $\Omega_{\rm GW}$ is unchanged under a rescaling of $V_0$. The only relevant consequence of changing $V_0$ is to change the frequency of the GW, such that with lower $V_0$ one can realize frequencies in the observable range.     
In the following we will discuss the models (\ref{eq:hilltop}) and (\ref{eq:tribrid}) on the same footing.

\begin{figure}[t]
\centering
\includegraphics[width=0.45\textwidth]{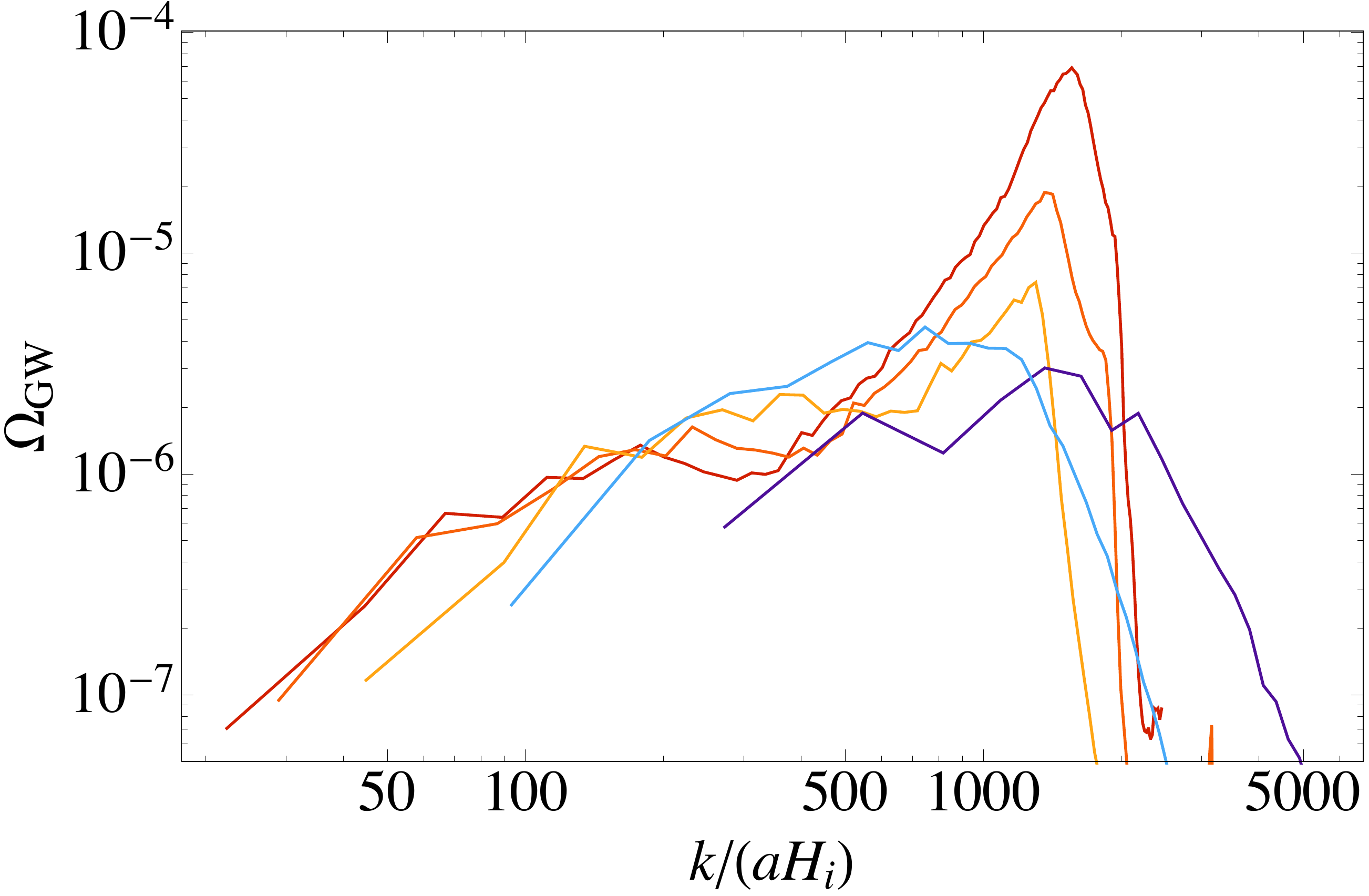}
\caption{Evolution of the spectrum of GW as a function of the physical momentum $k/(a H_i)$. The spectra are obtained from a lattice simulation of the model~(\ref{eq:hilltop}) with $128^3$ points. Table \ref{tab:IC_and_parameters} specifies the simulation setup. 
The lines correspond to the following times, scale factors:  
$t=573/m_\phi$, $a=1.47$ (purple);
$t=5544/m_\phi$, $a=4.29$ (blue);
$t=18924/m_\phi$, $a=8.9$ (yellow);  
$t=38040/m_\phi$, $a=13.81$ (orange); 
$t=57156/m_\phi$, $a=17.94$ (red).   
 }
\label{fig:spec}
\end{figure}

{\bf Results:}
Fig.~\ref{fig:spec} shows the evolution of the spectrum of GW in a lattice with $128^3$ points. We can distinguish three different stages: 
The initial stage corresponds to the linear growth of the spectrum (purple line in Fig.~\ref{fig:spec}), where GW are produced during tachyonic preheating and tachyonic oscillations, which are characteristic of hilltop inflation. 
Afterwards, in the second stage, the fluctuations become non-linear and oscillons form, resulting in a widening of the spectrum (blue line). 
Finally, the ``oscillon phase'' follows as third phase. As one can see from Fig.~\ref{fig:spec}, a peak in the GW spectrum forms and continues growing, becoming more and more significant (yellow to red lines).

It is during this third phase that our result for the GW spectrum differs strongly from the one of the previous study: While for the symmetric potential studied in \cite{Zhou:2013tsa} the growth stops and is followed by a phase of spherically symmetric oscillons where the production of GW is highly suppressed, in the models (\ref{eq:hilltop}) and (\ref{eq:tribrid}) the spectrum continues to grow and the amplitude of the peak becomes orders of magnitude larger than the 
linear preheating spectrum.

\begin{figure}[t]
\centering
\includegraphics[width=0.40\textwidth]{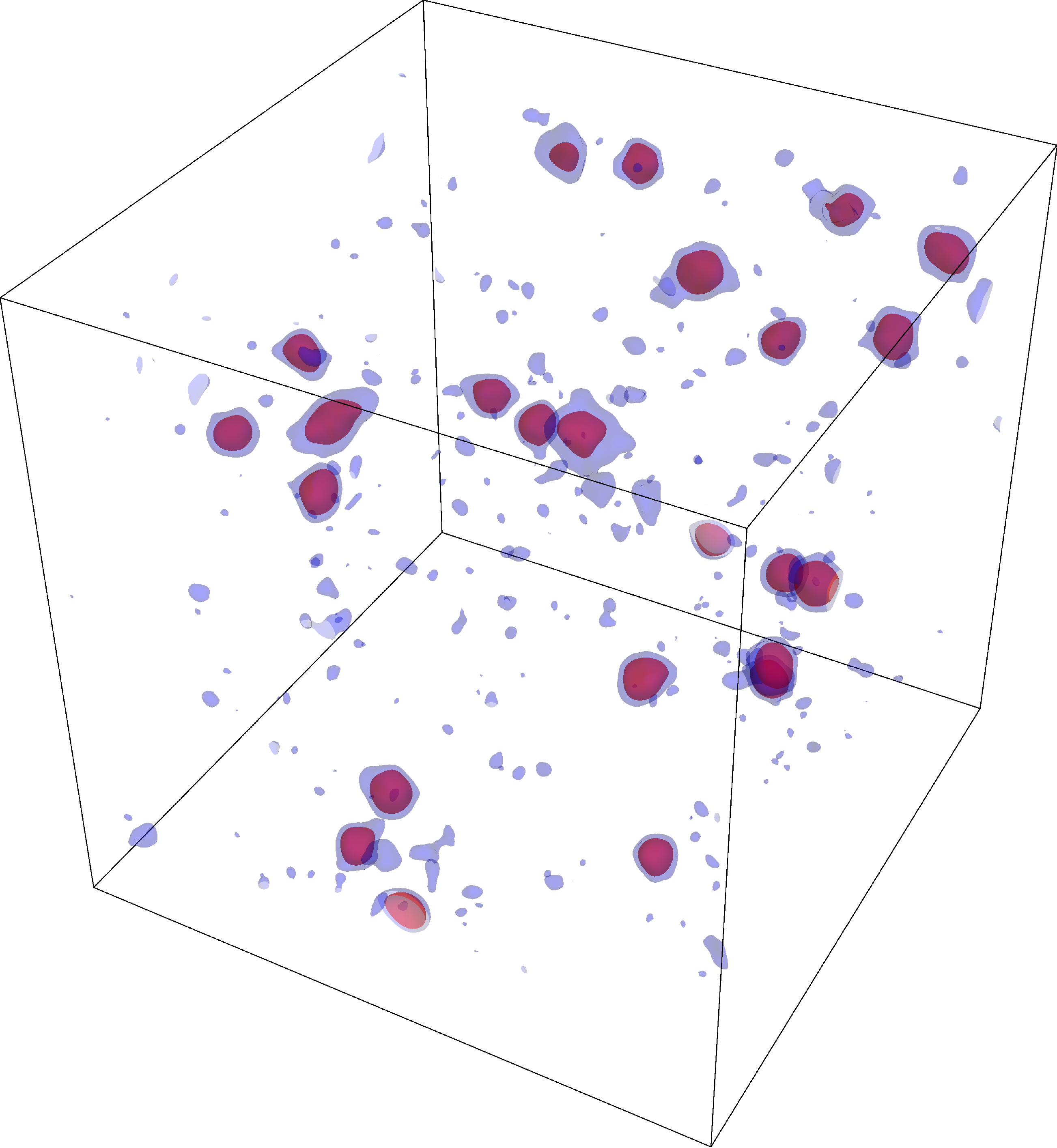}
\caption{Snapshot of the energy density during the ``oscillon phase'', at $a=5.35$, with energy density contours at $6\langle\rho\rangle_\mathcal{V}$ (blue) and $20\langle\rho\rangle_\mathcal{V}$ (red), obtained from a lattice simulation of model (\ref{eq:hilltop}) with $v=10^{-2} m_{\rm Pl}$, $V_0 = 10^{-19} m^4_{\rm Pl}$ and $p=6$. The lattice size is $L=0.02/H_{i}$ with $256^3$ points.    }
\label{fig:slice}
\end{figure}

Furthermore, we find that in contrast to lattice simulations of the symmetric model of \cite{Zhou:2013tsa}, where the oscillons form a nearly static network, in the asymmetric models they are not as isolated and collisions between the oscillons continue to happen until the end of the simulations ($\sim3$ $e$-folds after the oscillons have formed). Also, they do not become spherical as efficiently as in the symmetric model. In addition to the oscillons, which we illustrate in the time slice of energy overdensities shown in Fig.~\ref{fig:slice} as regions with  $\rho\geq20\langle\rho\rangle_\mathcal{V}$, other, less energetic overdensities  above $6\langle\rho\rangle_\mathcal{V}$ but below $20\langle\rho\rangle_\mathcal{V}$ are visible.

 Such additional overdensities are not spherically symmetric and may also contribute to the production of GW. We argue that the combination of these effects, originating from the considered generic class of asymmetric potentials, lead to a continuous growth of the GW spectrum during the ``oscillon phase''.    

Turning to the implications of the continuous growth of the GW spectrum, it is important to note that when we stopped our simulations $\sim3$ $e$-folds after the oscillons have formed, the oscillon peak in the GW spectrum was still growing, and there is no reason to assume that it would stop growing immediately after the end of the simulations. If the spectrum continues to grow, we may reach the point where a full general relativity simulation is necessary to obtain reliable results for the GW spectrum and also include the backreaction effects, which is beyond the scope of this letter. Such a large amplitude of GW may then also lead to constraints on inflation models of hilltop type from the BBN bound \cite{Shvartsman:1969mm,Maggiore:1999vm}, which requires $\Omega_{\rm GW,\,0}h^2 \lesssim 5\times10^{-6}$ from preheating.

\begin{figure}
\centering
\includegraphics[width=0.45\textwidth]{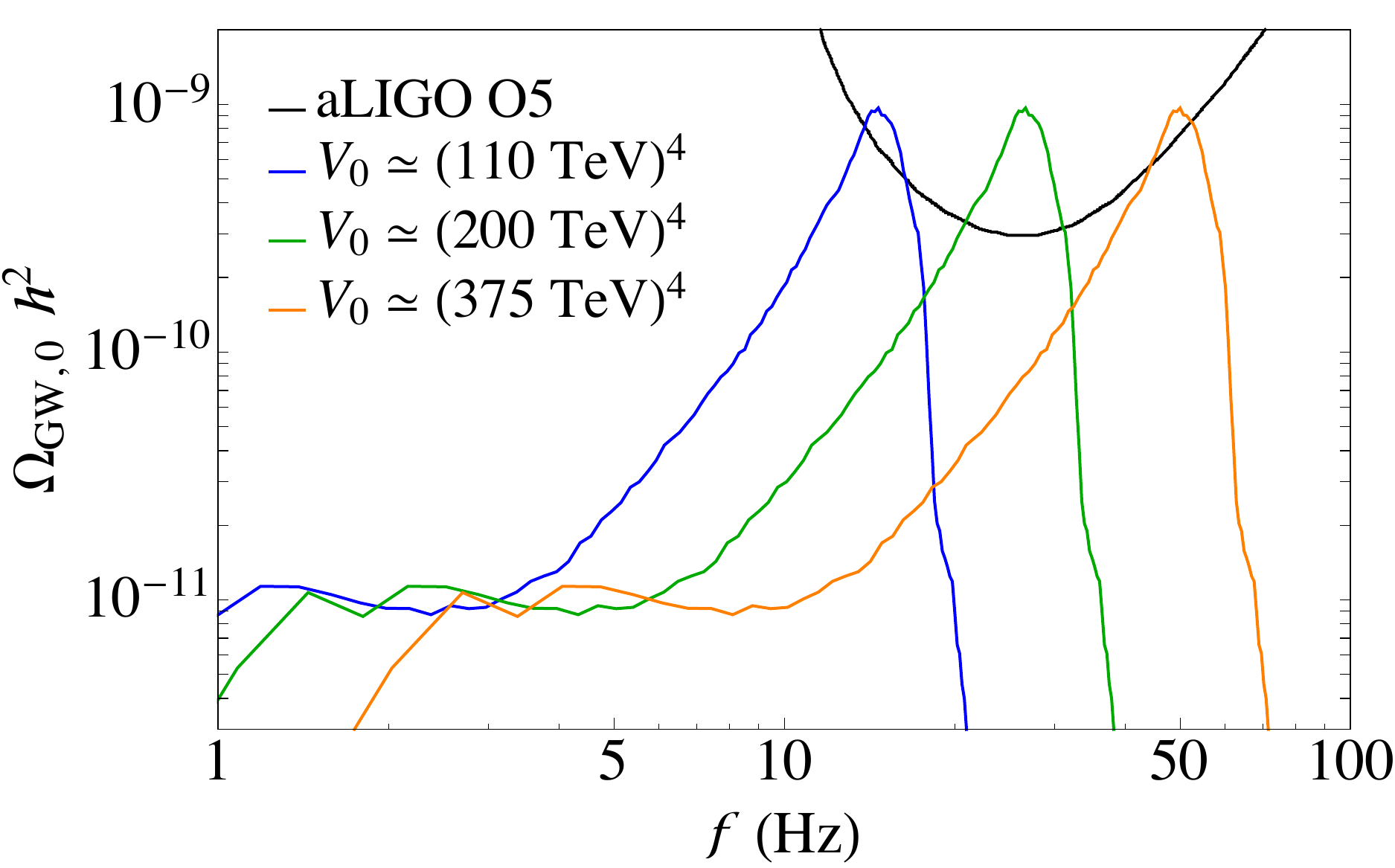}
\caption{Example predictions for gravitational wave spectra today, obtained from simulations of model~(\ref{eq:tribrid}) with parameters given in Table \ref{tab:IC_and_parameters} and using the results at $a_e = 15.51$. The spectra are shown for $V_0\simeq(110\,\textrm{TeV})^4$ (blue), $V_0\simeq(200\,\textrm{TeV})^4$ (green), $V_0\simeq(375\,\textrm{TeV})^4$ (orange) and compared to the expected sensitivity curve of the fifth observing run (O5) of the aLIGO--AdVirgo detector network \cite{TheLIGOScientific:2016wyq}.}
\label{fig:obs}
\end{figure}

Finally, let us discuss the prospects for observing the GW produced during the ``oscillon phase''. To this end, we consider the GW spectra from the end of our simulation, which gives a conservative estimate for the produced GW. We expect the peak from the oscillons to continue growing, which would further improve the detectability. As discussed above, observation prospects are greatly enhanced if we consider models of the form of Eq.~(\ref{eq:tribrid}), where $V_0$ is not constrained by the amplitude of the CMB temperature fluctuations. Fig.~\ref{fig:obs} shows the GW spectra obtained from a lattice simulation of the  model of Eq.~(\ref{eq:tribrid}) with simulation setup given in Table \ref{tab:IC_and_parameters}. Since the GW amplitude remains unchanged when changing $V_0$, to produce the plot in Fig.~\ref{fig:obs} we simply rescaled the frequency according to Eq.~(\ref{eq:fGW}), assuming reheating ends at $t_e$ (i.e.\ $a_e/a_* = 1$). For example, setting $V_0\simeq4.8\times10^{-53}m_{\rm Pl}^4\simeq(200\, {\rm TeV})^4$ leads to a frequency of $f\sim 30\,{\rm Hz}$, while the amplitude remains unchanged and the peak lies above the  expected sensitivity curve of the aLIGO--AdVirgo detector network which is expected for the fifth observing run (O5) \cite{TheLIGOScientific:2016wyq}. The planned Einstein Telescope detector \cite{ET} would have an even lower sensitivity (for more details see e.g.\ \cite{Sathyaprakash:2009xs,Moore:2014lga}).

We note that Fig.~\ref{fig:obs} is just an example, and indeed various parameters can affect the GW spectrum. First of all, if reheating continues after the end of the simulation at $t_e$ until $t_*>t_e$ with equation of state $w=0$, both the frequency and the amplitude are stretched to lower values, with $f\propto (a_e/a_*)^{1/4}$ and $\Omega_{{\rm GW},\,0}\propto a_e/a_*$. For $V_0=\mathcal{O}(100\, {\rm TeV})^4$, this can push the peak of the GW spectrum in the sensitivity region of the BBO \cite{BBO} and DECIGO \cite{DECIGO} experiments. 
Also, changing the vacuum expectation value $v$ of $\phi$ would affect the GW spectrum, since the scale of the peak is proportional to the mass of the oscillating field, which is inversely proportional to $v$. Thus, smaller $v$ would lead to larger $f$. Finally, the parameter $p$ sets the degree of asymmetry of the potential around the minimum: Larger $p$ means a potential which is steeper for $|\phi|>v$ and flatter for $|\phi|<v$. We expect this to affect the production of GW. 

In summary, we found that the gravitational wave production from the oscillons in the considered class of asymmetric potentials does not stop after the oscillon formation phase but leads to a continuous growth of the gravitational wave spectrum at a characteristic peak frequency, with an amplitude orders of magnitude above the spectrum from the initial phases of preheating. 
\begin{acknowledgements}
This work has been supported by the Swiss National Science Foundation. We thank Mar Bastero-Gil and David Nolde for useful discussions.
\end{acknowledgements}

\end{document}


\begin{center}

{\Large {\bf Erratum: Gravitational Waves from Oscillons after Inflation\\[1mm] [Phys. Rev. Lett. 118, 011303 (2017)]}}\\[8mm]

Stefan Antusch,
Francesco Cefal\`{a} and 
Stefano Orani

\end{center}

\vspace*{0.20cm}

\centerline{
Department of Physics, University of Basel,}
\centerline{\it
Klingelbergstr.\ 82, CH-4056 Basel, Switzerland}

\vspace*{1.2cm}
Although the general statement of the paper~\cite{Antusch:2016con} that asymmetric oscillons can produce a characteristic peak in the GW spectrum remains valid, we would like to correct the results of the lattice simulations for the example model we presented there. The reason for the incorrect results is that the accuracy of the finite-difference approximation for the spatial derivatives that we used has turned out to be insufficient to correctly describe the dynamics with $128^3$ lattice points and throughout the time period considered in~\cite{Antusch:2016con}. We have noted the failure of a convergence test in our paper~\cite{Antusch:2017vga}, and also a recent paper~\cite{Amin:2018xfe} has made this observation.

\begin{figure}[ht]
\centering
\subfigure{\includegraphics[width=0.45\textwidth]{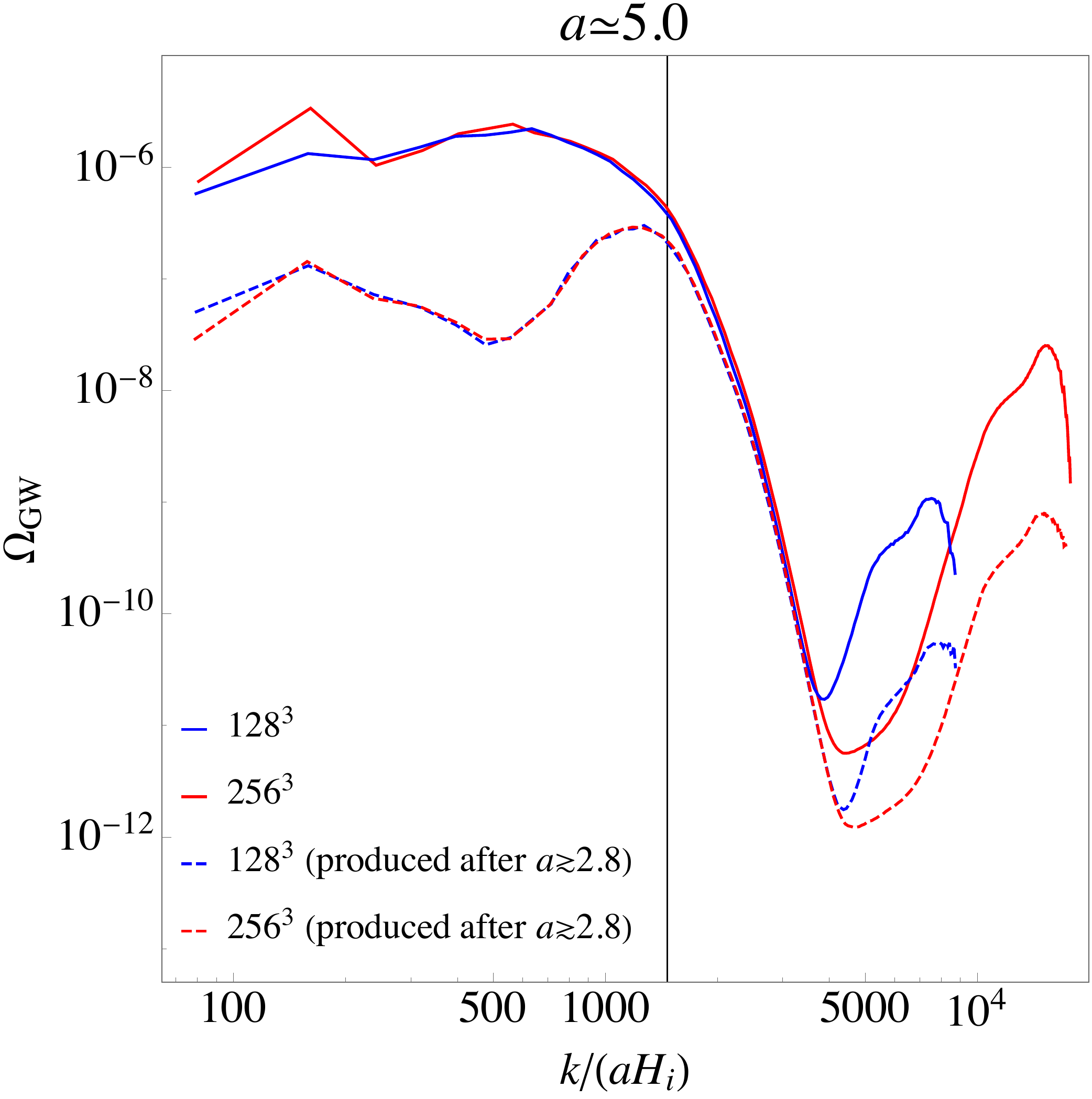}}
\hfill
\subfigure{\includegraphics[width=0.45\textwidth]{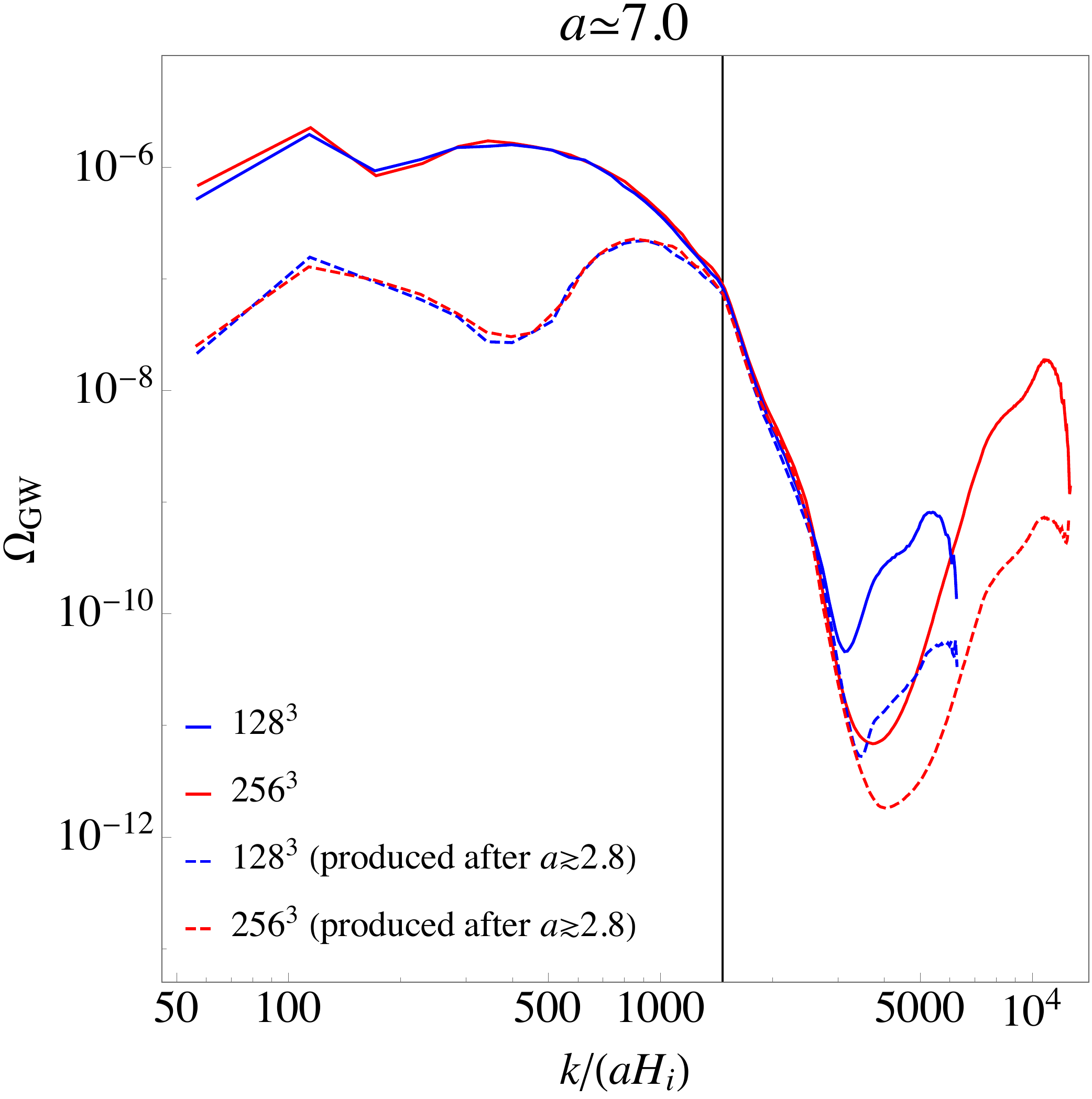}}
\hfill
\subfigure{\includegraphics[width=0.45\textwidth]{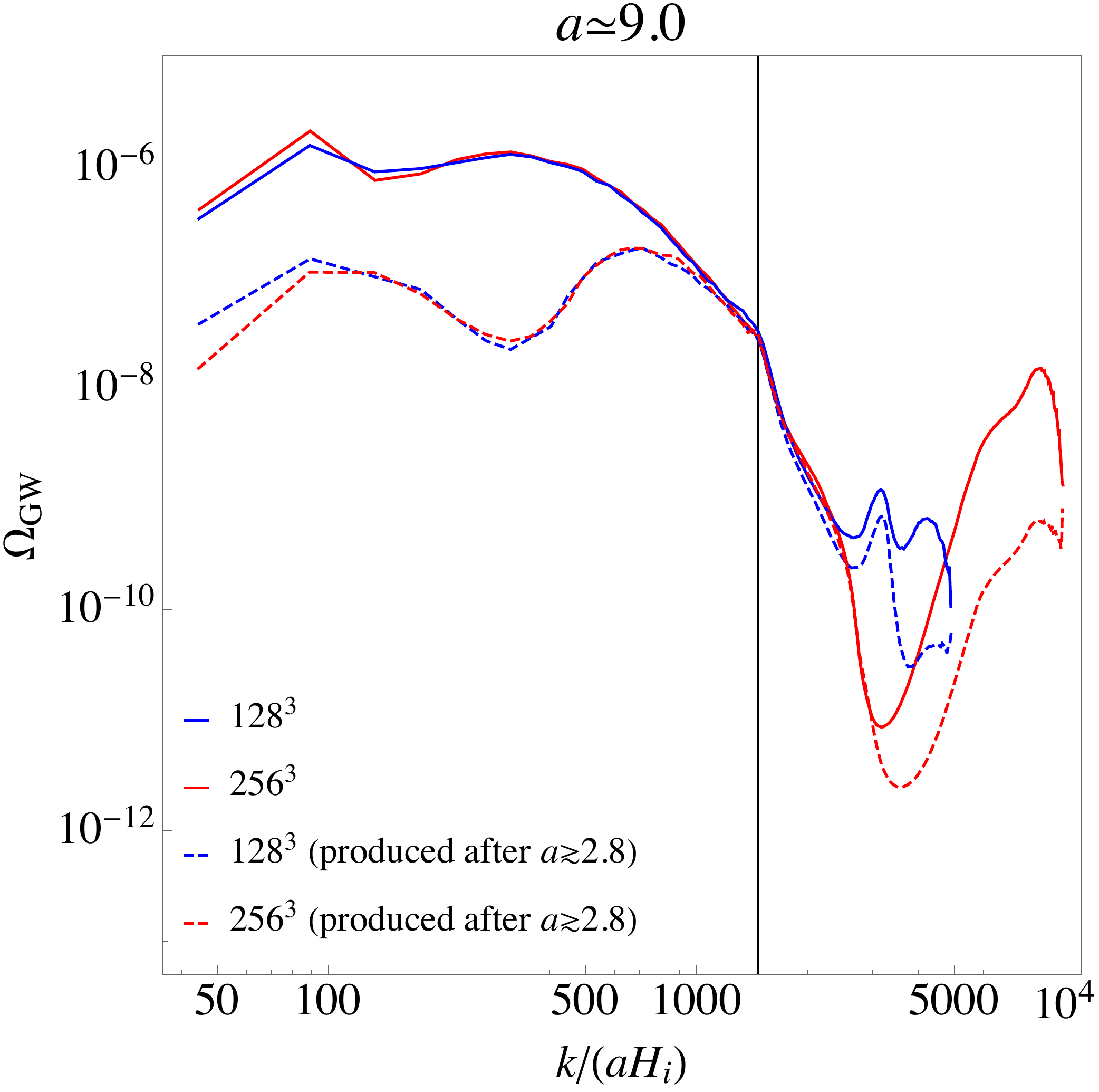}}
\hfill
\subfigure{\includegraphics[width=0.45\textwidth]{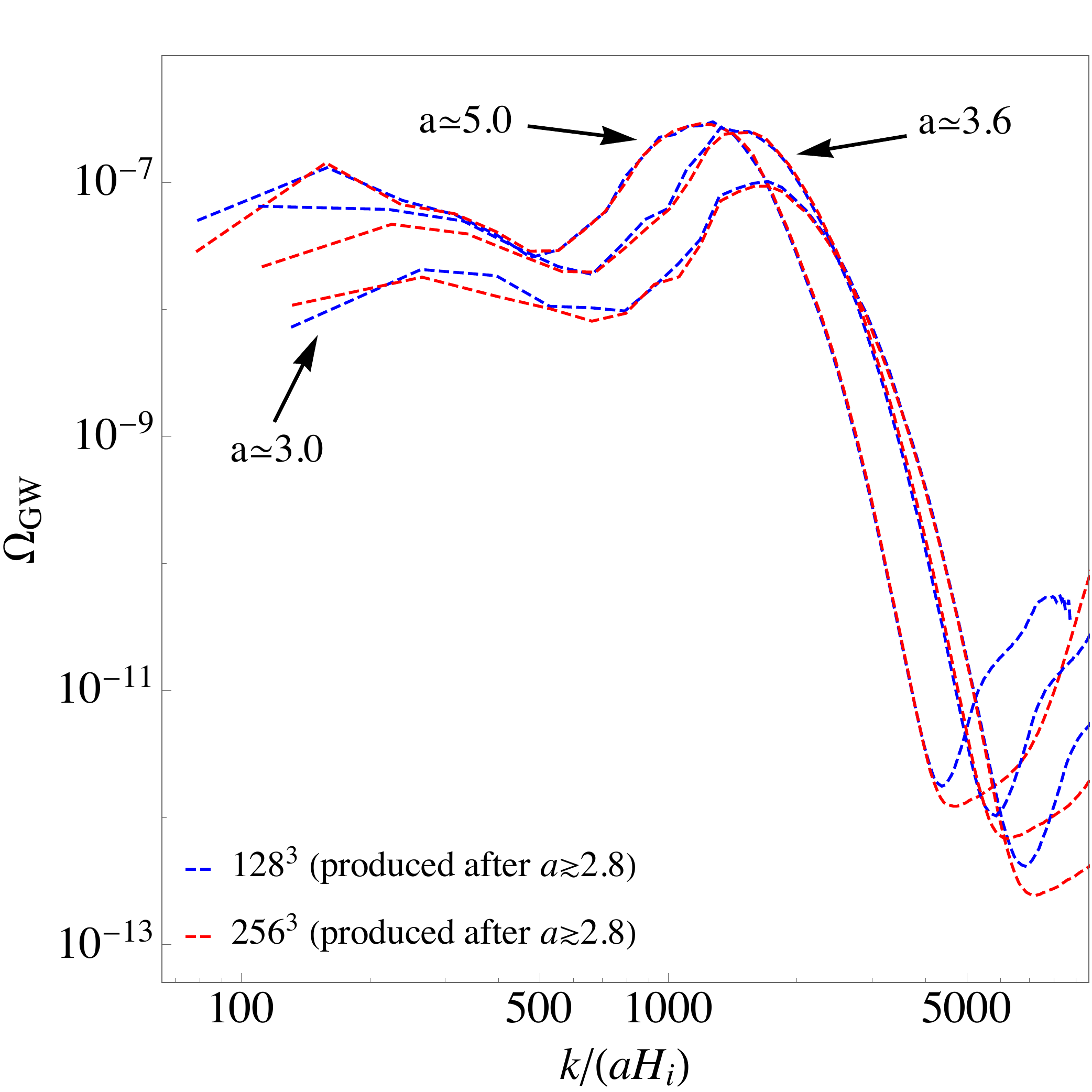}}
\caption{The plots show results from lattice simulations with $128^3$ lattice points (blue) and with $256^3$ points (red). The solid lines show the GW spectrum obtained from simulations which were performed as in~\cite{Antusch:2016con} but with more accurate approximations for the spatial derivatives as explained in the text. One can see that both simulations carried out with different resolutions agree very well, which indicates that the resolutions of both simulations are sufficient up to $a\simeq 9$. The dashed lines show the GW spectrum generated after $a = 2.8$, in order to distinguish the GW signal produced by oscillons from the foreground produced earlier from tachyonic oscillations. We can see that in both simulations a peak in the GW spectrum forms at a characteristic frequency corresponding to the physical mass scale (denoted by the solid black line). However, this growth stops at some point and the peak gets redshifted (cf.\ lower right plot). Although GWs continue to be produced at the characteristic frequency, the production efficiency decreases over time and the peak remains too low to be clearly visible, when compared to the foreground GW spectrum.} 
\label{fig:GW_specs}
\end{figure}

Correct results are now obtained with a more accurate approximation for the spatial derivatives, where we have replaced the second order scheme originally implemented in LatticeEasy \cite{Felder:2000hq}, with one that is fourth order accurate in space. The plots in Figure~\ref{fig:GW_specs} show the results of two lattice simulations (that were performed with exactly the same setup as in~\cite{Antusch:2016con}): one with $128^3$ points (solid blue) and one with $256^3$ points (solid red). We see that both simulations agree very well for the relevant scales, which indicates that the resolutions of both simulations are sufficient up to $a \simeq 9$.\footnote{We note that the features in the right tails of the GW spectra in Figure~\ref{fig:GW_specs} close to the cutoff, where the simulations apparently disagree, should be ignored, since they are known to be unphysical.} 

To distinguish the signal produced by asymmetric oscillons from that produced during preheating from tachyonic oscillations, we have performed additional simulations in which the GW produced during the early phase of preheating are ignored. In practice, this has been done by initializing the GWs after the end of the tachyonic oscillations at $a\simeq2.8$. 
The results of these simulations are also shown in Figure~\ref{fig:GW_specs}, corresponding to the dashed lines. One can see that a peak in the GW spectrum forms at a characteristic frequency (which corresponds to the physical mass scale). However, this growth stops at some point and the peak gets redshifted (cf.\ lower right plot in Figure~\ref{fig:GW_specs}). Although GWs continue to be produced at the characteristic frequency, the production efficiency decreases over time and the amplitude of the GW spectrum is not large enough to be visible compared to the foreground from the earlier produced GWs from the tachyonic oscillations. The example model analysed in our paper~\cite{Antusch:2016con} does therefore not feature a dominant peak in the GW spectrum due to the oscillons, which implies that the oscillon effects are unobservable at LIGO for this model.

On the other hand, in our recent paper~\cite{Antusch:2017flz}, we found that a pronounced oscillon peak in the GW spectrum can be produced from the overall K\"{a}hler modulus in the KKLT scenario (cf.\ Figure~9 in~\cite{Antusch:2017flz}). A successful convergence test has been made with simulations with $256^3$ and $512^3$ lattice points. For the parameter choices in~\cite{Antusch:2017flz} the frequency of the GWs is above the LIGO reach. The search for example models testable by future runs of LIGO is ongoing. 

